\documentclass[12pt,oneside]{article}
\usepackage{amsfonts,amssymb,graphicx}
\def\be{\begin{equation}}
\def\ee{\end{equation}}
\def\bea{\begin{eqnarray}}
\def\eea{\end{eqnarray}}
\def\ba{\begin{eqnarray*}}
\def\ea{\end{eqnarray*}}

\def\<{\langle}
\def\>{\rangle}
\def\~{\tilde}
\def\s{\sigma}

\def\a{\alpha}

\def\t{\tau}

\def\ds{\displaystyle}

\newcommand{\av}[1]{\mbox{{\rm Av}}\left[#1\right]}

\newtheorem{proposition}{Proposition}

\newtheorem{lemma}{Lemma}

\begin{document}
%
%
\begin{center}
\vspace{1truecm}
{\bf\sc\Large correlation inequalities for spin glass\\ in one dimension}\\
\vspace{1cm}
{Pierluigi Contucci $^{\dagger}$, \quad Francesco Unguendoli $^{*}$}\\
\vspace{.5cm}
{\small $^\dagger$ Dipartimento di Matematica, Universit\`a di Bologna, {e-mail: {\em contucci@dm.unibo.it}}} \\
\vspace{.5cm}
{\small $^{*}$ Dipartimento di Matematica, Universit\`a Modena, {e-mail: {\em unguendoli@unimore.it}}}\\
\end{center}
\vskip 1truecm
\begin{abstract}\noindent
We prove two inequalities for the direct and truncated correlation 
for the nearest-neighboor one-dimensional Edwards-Anderson model with symmetric quenched disorder. 
The second inequality has the opposite sign of the GKS inequality of type II. In the non symmetric case with positive average
we show that while the direct correlation keeps its sign the truncated one changes sign when crossing a suitable
line in the parameter space. That line separates the regions satisfying the GKS second inequality and the one proved here.
\end{abstract}

\section{Introduction and Results}

In a recent paper \cite{CL} a correlation inequality was proved 
for spin systems with quenched symmetric random interaction in arbitrary dimension, extending 
a previous result for the Gaussian case \cite{CG}. 
That inequality yield results for spin glasses similar to those obtained for ferromagnetic systems from
the first GKS inequality \cite{Gr,Gr2,KS} e.g. it gives monotonicity of the pressure 
in the volume and bounds on the surface pressure. 
Other inequalities were considered: in particular the extension to non-symmetric interactions and 
possible versions of a second type GKS inequality. 

In this work we study the $d=1$ case with nearest neighboor interaction. In the same spirit of the GKS
systems no assumption of translation invariance is made on the interaction distributions and by 
consequence our results cannot be obtained by an exact solution. We prove that both the 
inequality of the first type does extend to the non symmetric case and that an inequality 
of the second type holds indeed in the symmetric case. A similar result with a complete
proof of inequalities of type I and II has been obtained so far only in the Nishimori line \cite{CMN,MNC}.

Let us consider a chain with periodic boundary condition 
\ba 
H(\s, J) \; = \; -\sum_{i=1}^{N}J_{i}\s_i\s_{i+1} 
\ea with $\s_{N+1}=\s_1$. The
random variables $J_i$ have independent distributions $p(J_i)$. Those fulfills one of the three
following hypothesis, which will be called system I, II and III in the remaining part of the paper:\\
{\bf I)}
\ba
p(|J_i|) \geq p(-|J_i|), \quad \forall i \; \; and \; \; \forall |J_i| \in \mathbb{R}^{+}
\ea
{\bf II}) the $J_i$ are symmetric around a positive mean $\mu_i >0$:
\ba
p(\mu_i +|J_i|) = p(\mu_i - |J_i|), \quad \forall i \; \; and \; \; \forall |J_i| \in \mathbb{R}^{+}
\ea
In the case of discrete variables: $J_i=\mu_i\pm J^{(i)}$, $p(\mu_i+J^{(i)})=p(\mu_i-J^{(i)})=1/2$, we assume
that $J^{(i)} > \mu_i$ (see below for further explanations) and we introduce the notations:
\ba
& a_i = \mu_i + J^{(i)}& \\
& -b_i = \mu_i - J^{(i)}& \\
& a_i, b_i > 0 &
\ea
{\bf III}) the $J_i$ are discrete variables taking on values $\pm J^{(i)}$ with $J^{(i)}>0$ such that:
\ba
\textit{where}  \quad p_i = p(J^{(i)}), \;\;\; q_i = p(-J^{(i)}) \; ,  
\ea
and 
\ba
\a := {\ds \prod_i (p_i - q_i)} \; \ge 0 \; .
\ea
Let $\omega_{h}$ the thermal average of the quantity $\sigma_h\sigma_{h+1}$, $\omega_{h,k}$ 
that of the quantity $\sigma_h\sigma_{h+1}\sigma_k\sigma_{k+1}$ and $\av{\cdot}$ the average over
the quenched disorder.

Our main results are:
\begin{proposition}\label{pp1}
For all three systems:
\be\label{cu1}
\av{J_h \omega_h} > 0, \quad \quad \forall h=1...N
\ee
\end{proposition}

\begin{proposition}\label{pp2}
For systems I and III with $\a=0$:
\be\label{g2}
\av{J_h J_k (\omega_{hk} - \omega_h \omega_k)} < 0, \quad \quad \forall h, k = 1...N, \quad h \neq k
\ee
\end{proposition}

\begin{proposition}\label{pp3}
For system III, with $\a >0$, 
the following properties hold:
\begin{itemize}
\item[ ] $\forall l$, there exists in the  $(J^{(l)} , \a)$ quadrant, a curve $\a(J^{(l)})$
such that the quantity
\be
\av{J_h J_k (\omega_{hk} - \omega_h \omega_k)}
\ee
changes its sign from negative to positive when crossing the curve  $\a(J^{(l)})$ by increasing $\alpha$ 
and such that on the curve $\a(J^{(l)})$
\be
\av{J_h J_k (\omega_{hk} - \omega_h \omega_k)} = 0, \quad \quad \forall h, k = 1...N, \quad h \neq k \; .
\ee
Moreover $\av{J_h J_k (\omega_{hk} - \omega_h \omega_k)}$ is increasing in $\a$ along the $J^{(l)} = \textrm{const}$ lines.
\end{itemize}
\end{proposition}

\section{Proofs}
We start by proving the following lemmata.
\begin{lemma}\label{lem1}
System III can be rewritten as:
\be\label{pippe}
H(\t, K) = -K_N\t_N\t_{N-1} - \sum_{i=1}^{N-1}J^{(i)}\t_i\t_{i+1}
\ee
with:
\bea
K_N = J^{(N)}\prod_{i=1}^{N} {\rm sgn}(J_i) = \pm J^{(N)} \; . 
\eea
Setting $P = {\rm prob}(K_N=J^{(N)})$ and $Q= {\rm prob}(K_N=-J^{(N)})$ we have:
\bea
P&=& \frac{1+\prod_i (p_i - q_i)}{2} \\
Q&=& \frac{1-\prod_i (p_i - q_i)}{2}
\eea
\end{lemma}

\begin{lemma}\label{lem2}
Consider system II with discrete variables and assume that $\mu_h=0$ for at least one $h$.
Such a system can be rewritten as:
\ba
H(\t, K)=-\sum_{i=1}^{N}K_{i}\tau_i\tau_{i+1}
\ea
where:
\bea
K_h = J_h &=& \pm a_h, \quad a_h>0 \\
K_i &=& \left\{ \begin{array}{c} a_i > 0 \\ b_i > 0 \end{array} \right.
\eea
the two cases having probability $1/2$.
\end{lemma}
{\bf Proof of Lemma 2.1} \\
The proof is based on the Gauge transformation $\alpha_j=\prod_{1\le i < j}{\rm sgn}(J_i)$,
for $2\le j\le N$ $\a_1=1$. Set $\t_i=\a_i\s_i$ $H$ is given by (\ref{pippe}) with $K_N = J^{(N)}\prod_{i=1}^{N} {\rm sgn}(J_i)$.
We have now to compute the new probability measure for $\prod_{i=1}^{N} {\rm sgn}(J_i)$.
Clearly the expectation 
\ba
\av{\prod_{i=1}^{N} {\rm sgn}(J_i)}=\prod_{i=1}^{N} \av{{\rm sgn}(J_i)}=\prod_{i=1}^{N}(p_i-q_i)=P-Q \; .
\ea

\noindent
{\bf Proof of Lemma 2.2} \\
Group the bond configurations in couples that only differ for the sign of
$J_h$ and Gauge transform them using the same transformation of Lemma 3.1
What we obtain is:
\ba
K_i &=& |J_i | > 0 \\
K_h &=& J_h = \pm J^{(h)}
\ea
Moreover, since $p(K^{(h)}) = p(J^{(h)})$ or $p(K^{(h)}) = p(-J^{(h)})$,
$p(K^{(l)}) = 1/2$ for all $l$. \\

\noindent
Introduce, for system {\bf III}, the following shorthand notations:
\ba
C_i := \cosh(K^{(i)}) = \cosh(J^{(i)}) ; \quad \quad S_i := \sinh(K^{(i)}) = \sinh(J^{(i)})
\ea
\\ \noindent
{\bf Proof of Proposition 1.1} \\ First we prove the thesis for discrete variables (system II and III).
The partition function and correlation of an $N$ spins chain with periodic boundary conditions
can be written as:
\bea
& Z = {\ds \prod_i} C_i + {\ds \prod_i} S_i & \\
& \omega_h = {\ds \frac{1}{Z}} \left[ S_h \, {\ds\prod_{i \neq h} }C_i + C_h \, {\ds\prod_{i \neq h} }S_i \right]
\eea
{\bf System III} Using Lemma 2.1 one has:
\ba
{\rm Av}_{\{ J \} }[J_h \omega_h] &=& {\rm Av}_{(K_h)}[K_h \omega_h] = K^{(h)}\left\{ P\omega |_{k_h=k^{(h)}}-Q\omega |_{k_h=-k^{(h)}} \right\} = \\ \\
&=& K^{(h)}\left\{ Q\left[ \omega |_{k_h=k^{(h)}}-\omega |_{k_h=-k^{(h)}}\right]+(P-Q) \omega |_{k_h=k^{(h)}} \right\} \ge 0
\ea
due to the first Griffith's inequality for ferromagnetic systems.
\\ \noindent
{\bf System II} (discrete variables) Since the pressure is a convex function of the $\mu_i$'s we can prove our theorem
for $\mu_h = 0$. If for some $i$ $J^{(i)}\le \mu_i$ the variable $J_i$ takes positives values and it doesnt influence the sign
of the average. Now, using lemma 2.2 and observing that $P=Q=1/2$ and the $J$ average is a linear combination
of $K_h$ average with all positive remaining $K_i$, we have the thesis with the same steps as before.
The extension to the continuos case is obtained by the usual method of integrating over
the positive parts of the $J_i$ distributions.

\noindent
{\bf Proof of Proposition 1.2 }\\
For discrete variables (system III), using the standard hyperbolic expansion:
\ba
\omega_{hk}  = {\ds \frac{1}{Z}} \left[ S_h S_k  {\ds\prod_{i \neq h,k} }C_i + C_h C_k {\ds\prod_{i \neq h,k} }S_i \right] \Rightarrow
\ea
\ba
\omega_{hk} - \omega_h \omega_k &=& \frac{1}{Z^2} \left\{ \left( S_h S_k  {\ds\prod_{i \neq h,k} }C_i + C_h C_k {\ds\prod_{i \neq h,k} }S_i \right)
                            \left({\ds \prod_i} C_i + {\ds \prod_i} S_i \right) + \right. \\ \\
&&      \quad \quad \quad  \left. - \left(S_h  {\ds\prod_{i \neq h} }C_i + C_h {\ds\prod_{i \neq h} }S_i \right)
                                \left(S_k {\ds\prod_{i \neq k} }C_i + C_k {\ds\prod_{i \neq k} }S_i \right) \right\} = \\ \\
&=& \frac{1}{Z^2} \left\{  {\ds\prod_{i \neq h,k} }(C_i S_i) \cdot \left( C_h^2 C_k^2 + S_h^2 S_k^2 - C_h^2 S_k^2 - S_h^2 C_k^2 \right) \right\} = \\ \\
&=& \frac{ {\ds\prod_{i \neq h,k} }C_i S_i }{\left( {\ds \prod_i} C_i + {\ds \prod_i} S_i \right)^2 }
\ea
If at least one of the random variables is symmetric we have: $P=Q=1/2$; using lemma 2.1 one has:
\ba
{\rm Av}[K_h K_k (\omega_{hk} -  \omega_h \omega_k)] &=& J^{(k)}  \cdot
                                    {\rm Av}_{(K_N)}\left[ \frac{K_h\cdot {\ds\prod_{i \neq h,k} }(C_i S_i)}{ (  \prod_i C_i +  \prod_i S_i )^2 } \right] = \\
&=& J^{(k)} J^{(h)} {\ds\prod_{i \neq h,k} }(C_i S_i) \cdot \frac{1}{2} \left\{ \frac{1}{(  \prod_i C_i +  \prod_i S_i )^2 } -
                                        \frac{1}{(  \prod_i C_i -  \prod_i S_i )^2 } \right\} = \\
&=& - 2 \,  J^{(k)} J^{(h)} {\ds\prod_{i \neq h,k} }(C_i S_i)\cdot \frac{\prod_i (C_i S_i) }{ (\prod_i C_i^2  - \prod_i S_i^2)^2} < 0
\ea
The extension to the continuous case is as above.

\noindent
{\bf Proof of Proposition 1.3 }\\
Let $\a > 0$ or equivalently $P = \frac{1+\a }{2} > \frac{1}{2} > Q = \frac{1-\a }{2}$. \par
We obtain analogously as before:
\ba
&{\rm Av}[K_h K_k (\omega_{hk} -  \omega_h \omega_k)] =  J^{(k)} J^{(h)} {\ds\prod_{i \neq h,k} }(C_i S_i)
                                {\ds \frac{P (\prod C_i - \prod S_i)^2 - Q  (\prod C_i + \prod S_i)^2 }{ (\prod C_i^2  - \prod S_i^2)^2}} =& \\
&={\ds \frac{ J^{(k)} J^{(h)} \prod_{i \neq h,k }(C_i S_i) }{(\prod_i C_i^2  - \prod_i S_i^2)^2}} \cdot \left\{ (P-Q) (\prod_i C_i^2 + \prod_i S_i^2) -
                                        2 \prod_i (C_i S_i) \right\}&
\ea
The sign of the previous expression is, by inspection, the same as that of the curly parentheses:
\ba
g(\a; \{ J \}) := \a  (\prod_i C_i^2 + \prod_i S_i^2) - 2 \prod_i (C_i S_i)
\ea
One obtains:
\begin{itemize}
\item[-] $\a = 0$ (zero mean spin glass)  $\Rightarrow g(\a; \{ J \}) < 0$;
\item[-] $\a = 1$ (ferromagnetic)  $\Rightarrow g(\a; \{ J \}) =  (\prod_i C_i - \prod_i S_i)^2 > 0$;
\item[-] for all $J^{(l)}$, $g(\a; \{ J \})$ is increasing function of $\a $;
\item[-] ${\rm Av}_{(K_h)}[K_h K_k (\omega_{hk} -  \omega_h \omega_k)] = 0$ on the $(J^{(l)}, \a)$ plane curve with $J^{(l)} >0$ and \linebreak $0 \leq \a \leq 1$
defined by:
\be
\a (J^{(l)}) = \frac{2 C_l S_l \prod_{i \neq l} (C_i S_i)}{ C_l^2 \prod_{i \neq l} C_i^2 + S_l^2 \prod_{i \neq l} S_i^2}
\ee
\end{itemize}

The proof of the inequalities for one dimensional systems with free
boundary conditions or for tree-like lattices is trivial since, due to the absence of loops
the partition function factorizes
\ba
{\bf Z} = 2^N \prod_i \cosh(\lambda_i J_i)
\ea
and by consequence the first inequality is fulfilled even without taking the average
and the second inequality reduces obviously to the equality to zero. \\

\section{Comments}
We proved that a one dimensional spin glass system fulfills a family of correlation inequalities 
without the assumption of translation invariance for the interaction distribution. The first inequality 
extends a similar one proved in \cite{CL} for any lattice and any interaction with zero mean value.
Here we have shown that the inequality is stable by suitable deformations of the zero mean hypotheses.
The inequality of type II proved here shows that in the zero mean case the truncated correlation function
has the opposite sign of the standard GKS inequality i.e. the case of interactions with zero variance and
positive mean. We have moreover identifyed the line crossing which the truncated correlation changes its sign.  
It would be interesting to establish if an inequality of type (\ref{g2}) is fulfilled
also in higher dimensions (see \cite{KNA}). In fact, as a straightforward computation shows in the Gaussian
case, if such an inequality holds then the overlap expectation would be monotonic
in the volume and several nice regularity properties would follow \cite{CG2}. We also mention that
the inequality (\ref{g2}) doesn't hold in general topologies as it was shown to us by Hal Tasaki
for a Bernoulli spin chain with an extra bond connecting two non adjacent sites. Moreover
a similar violation for the inequality (\ref{g2}) can be obtained in the case in which the disorder, 
still having zero average, is non symmetric.\\

{\bf Acknowledgments}. The authors thank Joel Lebowitz for many fruitful suggestions.
P.C. thank Cristian Giardina, Sandro Graffi, Frank Den Hollander and Hidetoshi Nishimori for useful discussions.

\end{document}